# FODS HADRON CALORIMETERS

A. Yu. Kalinin, A. V. Korablev, A. N. Krinitsyn, V. I. Kryshkin, V. V. Skvortsov, V. V. Talov,
L. K. Turchanovich, A. A.Volkov

(*Institute for high energy physics, Protvino*)

**Abstract**

There is described the design of the FODS hadron calorimeters, the procedure of the calibration and the results of the calorimeter study with electron and pion beams.

## Introduction

To study high $p_T$ particle production at 70 GeV hadron - hadron collision there was constructed double arm spectrometer FODS [1] which used hadron calorimeters for a trigger. During the spectrometer operational period the light yield of the scintillator and light shifting fibers efficiency dropped below acceptable level. Therefore the active calorimeter elements were replaced using the existent absorber.

## 1. Calorimeter design

The hadron calorimeter absorber (see fig. 1) consists of 38 rolled 25 mm thick iron plates welded on top and below to iron plates to fix the position. The gap between the plates is 9 mm.

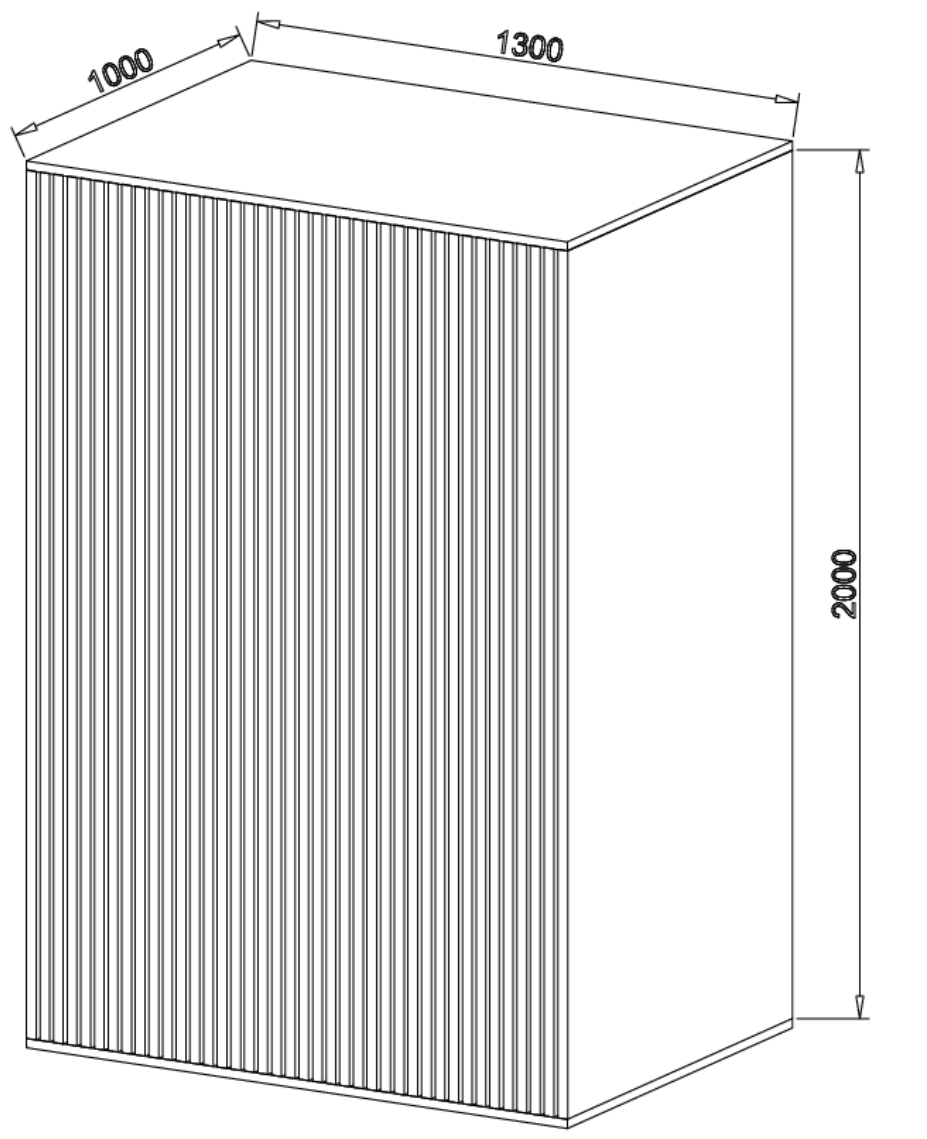

Fig. 1. Design of the hadron calorimeter absorber.

One side of the absorber is closed with a thin metal plate and on the other side there is a double wing door. The 16 tons absorber is installed on a platform with an electrical motor that moves the calorimeter on the rails.

The original design of the active elements was based on principle outlined in [2] and made provision for summed signal for a trigger and simultaneously provided for separate tower signals to measure particle coordinate and energy.

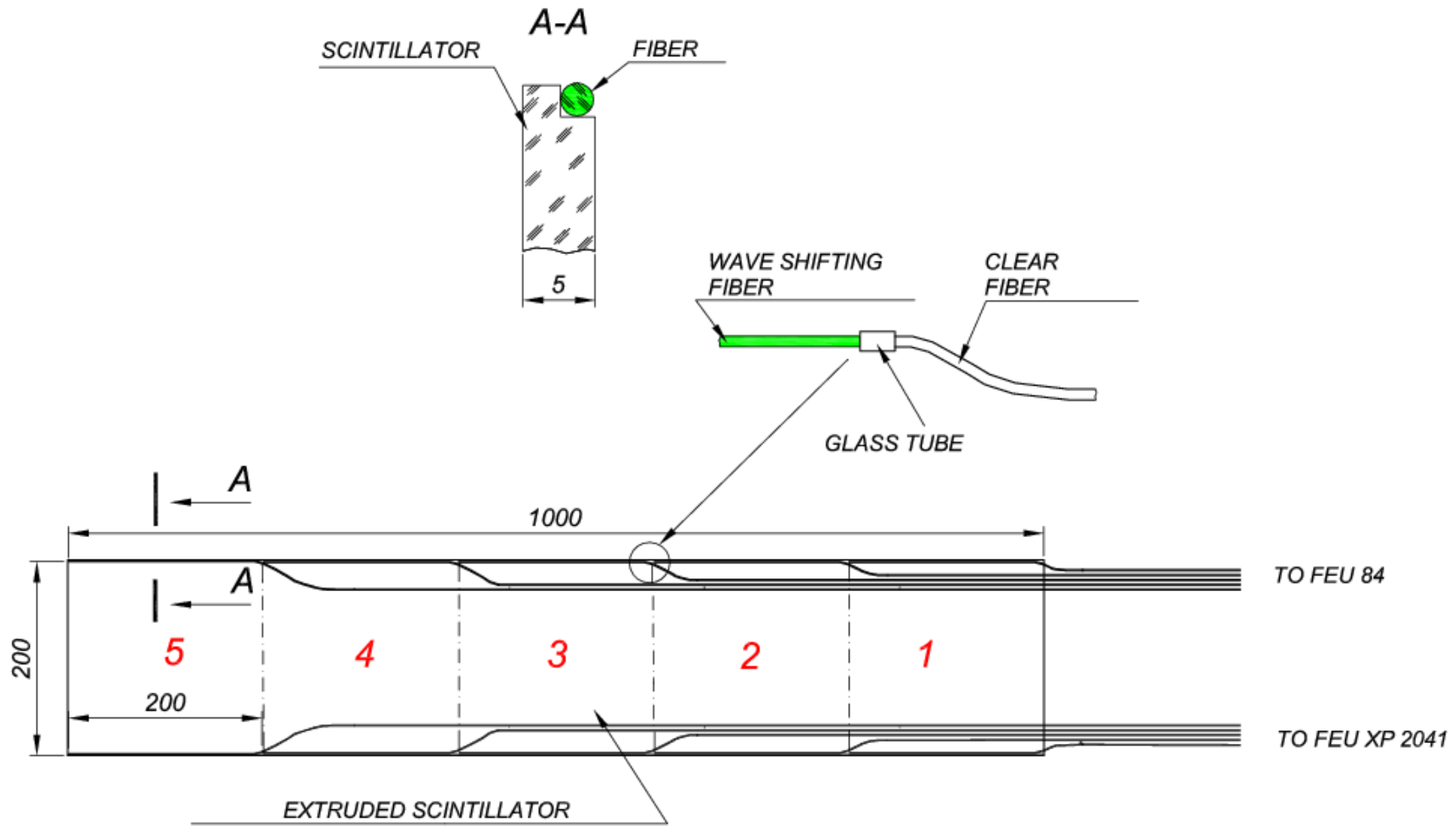

Fig. 2. The original design of the hadron calorimeter active element.

As fig. 2 shows an extruded scintillator 100 cm long had ledges along the both sides. A light shifting fiber 20 cm long and 1 mm diameter was stacked in the ledge. One fiber end was sputtered with aluminum and the other one was glued to a clear fiber inside of a glass tube. The scintillator was wrapped into aluminized Mylar. The clear fibres from one side of the scintillator went to a corresponding tower (5 towers in horizontal direction and 10 rows in height - 50 towers altogether) with FEU 84-3. All fibres from the other side of the scintillator were combined on the photocathode of XP 2041 to produce fast signal corresponding to the total energy release. This side had 5 fibers to improve light collection uniformity.

Fig. 3 shows routing of the fibers for one layer. Two calorimeters required 15 km of the polymer optical fibers that were produced by the research institute "Polimer".

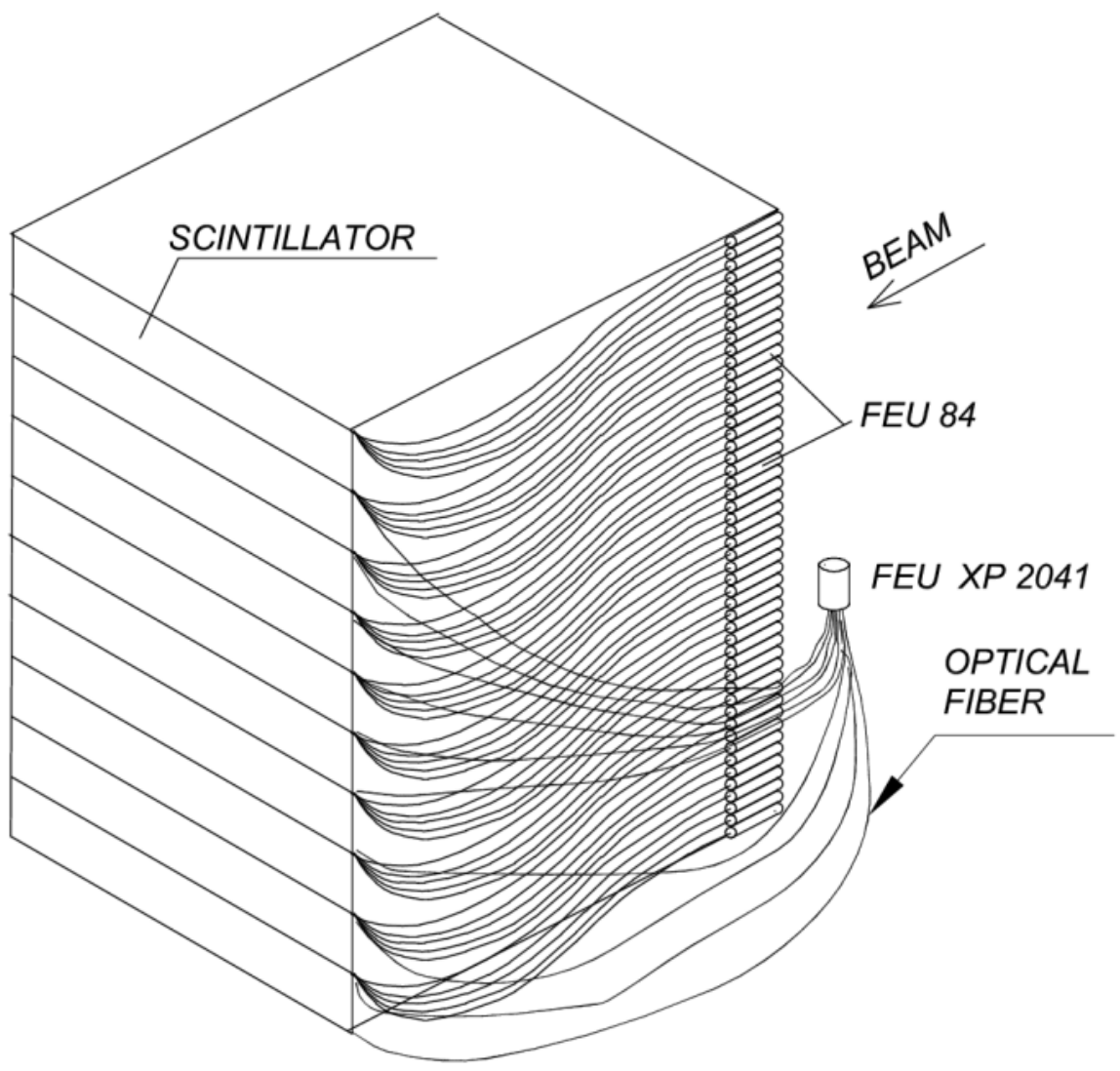

Fig. 3.The layout of the hadron calorimeter. The optical fibers are shown only for one layer.

It turned out that for the reconstruction of the FODS events the information from the calorimeter towers did not add anything to the information obtained from the measurements of particle trajectory in magnetic field. Therefore the new upgraded active elements based on a casting scintillator and high transparency and light shifting efficiency fiber do not have tower division to minimize the cost.

The new active element presented in fig. 4 consists of 5 scintillators with dimensions 192.5 × 194 × 3.4 mm$^3$. In the middle of the scintillator there is a groove1.2 mm deep, 1.2 mm wide. Its bottom has a cylindrical cross section. The fiber with wave length shifter Y11 [3] 1 mm diameter and 2.3 m length is inserted into the groove. The fibre ends were machined by diamond mill and one end was aluminized. The scintillators with the fiber were wrapped into reflective cover Tyvek [4] that also used as an additional rigid supporting structure.

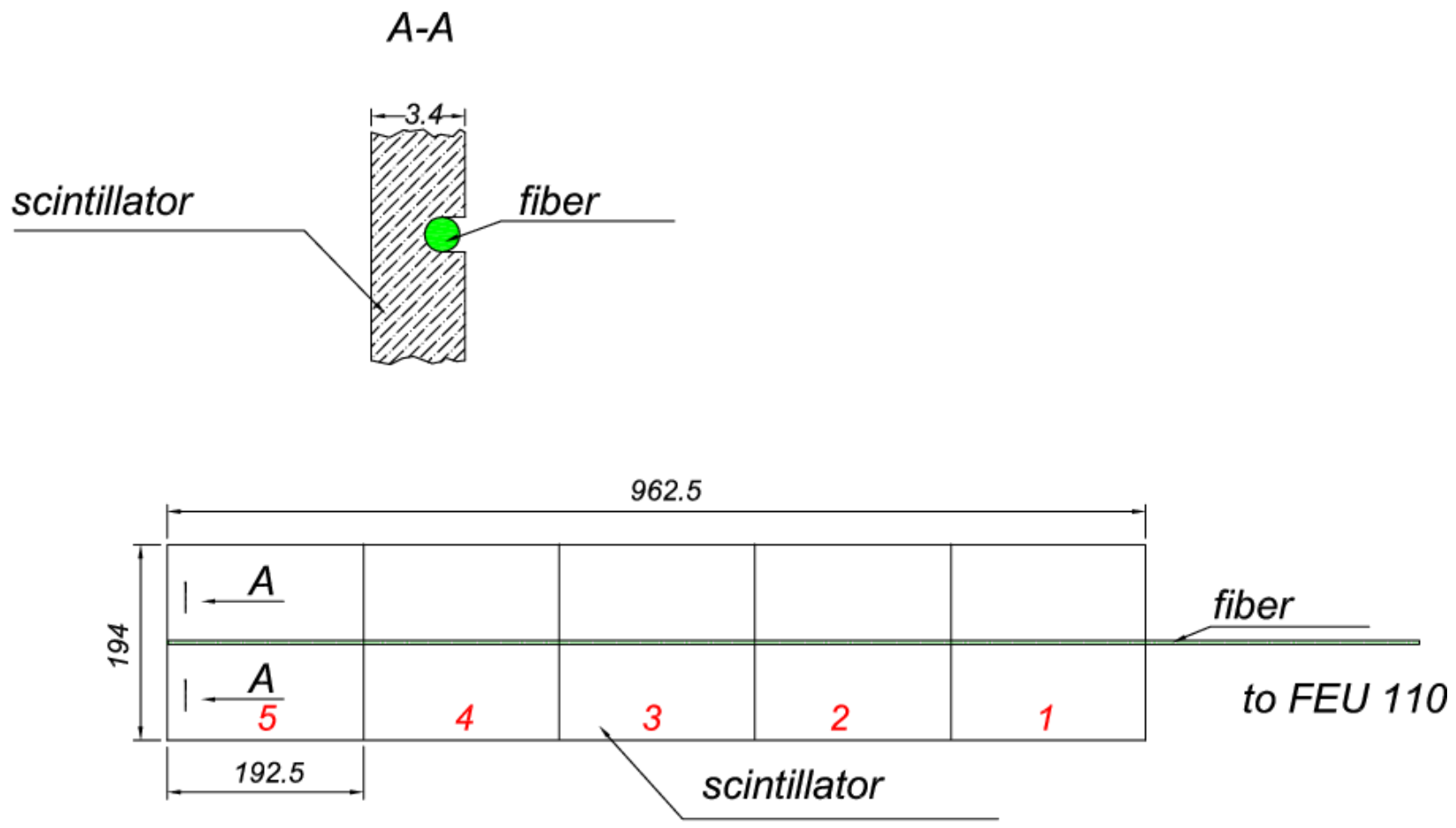

Fig. 4. The new hadron calorimeter active element.

The light collection uniformity of the elements was studied with a collimated radioactive source measuring current by FEU -85 contacted to fibre . Fig. 5a shows the dependence of current on the radioactive source position in the transverse direction. To improve the light collection uniformity different options of correction were studied. The red circles are correspond to the uniformity without any correction. The blue squares shows the uniformity of the light collection with a black tape 4 mm wide glued under the groove. The uniformity for the adopted solution is presented by black triangles due to simplicity and time stability. In this case under the groove a black tape was glued to Tyvek. To improve the longitudinal uniformity the tape width was increasing from distant end to the close to phototube end from 3 mm to 6 mm.

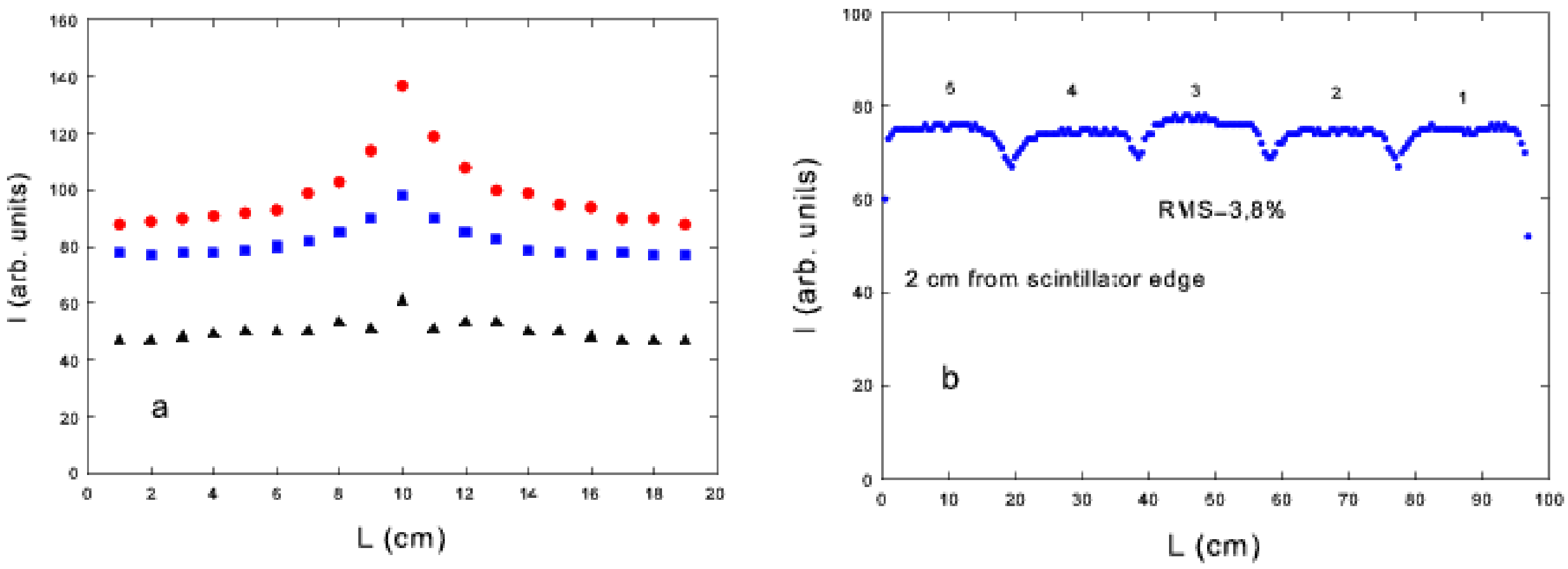

Fig. 5. The light yield of the calorimeter active element in the transverse direction a) and in longitudinal direction b).

The final correction of the longitudinal uniformity was made by gluing a black tape to Tyvek opposite to the narrow side of the scintillators (0.3 mm wide for number 2 and 0.5 mm wide for the scintillator number 1). The Fig. 5b presents the longitudinal light collection uniformity for the adopted solution. The drop of the signal between the scintillators is very hard to correct. The optical element light uniformity is about 7% (r.m.s.).

At IHEP there was produced a mould for injection molding and manufactured 3000 scintillators. The assembled active elements were studied on the test bench with radioactive source. The assembled 560 elements were divided into 2 groups and inserted into the calorimeters (fig. 6). The fibers in each calorimeter were collected into a bundle 18 mm diameter and coupled to the part of the photocathode of FEU -110 where the sensitivity was not less than 95% of the maximal one. The anode photomultiplier signal was fed to linier splitter and then to ADC and a discriminator to generate a trigger on a particle energy above a set threshold.

Such design has high operating speed and simple calibration scheme. To control the calorimeters performance the light of a green LED was send through an optical fiber to the photomultiplier photocathode.

Only 28 gaps out of 37 were filled with the active elements that corresponded to 4 nuclear absorption lengths and 95% energy absorption of hadron with momentum 10 GeV/c. The final volume nonuniformity of the calorimeters did not exceed 10 %. The energy resolution of the calorimeters is not an important factor because in front of the calorimeters the spectrometers of Cherenkov radiation containing rather much materials (0.2-0.6 nuclear absorption lengths in the aperture of 1 m diameter) are installed.

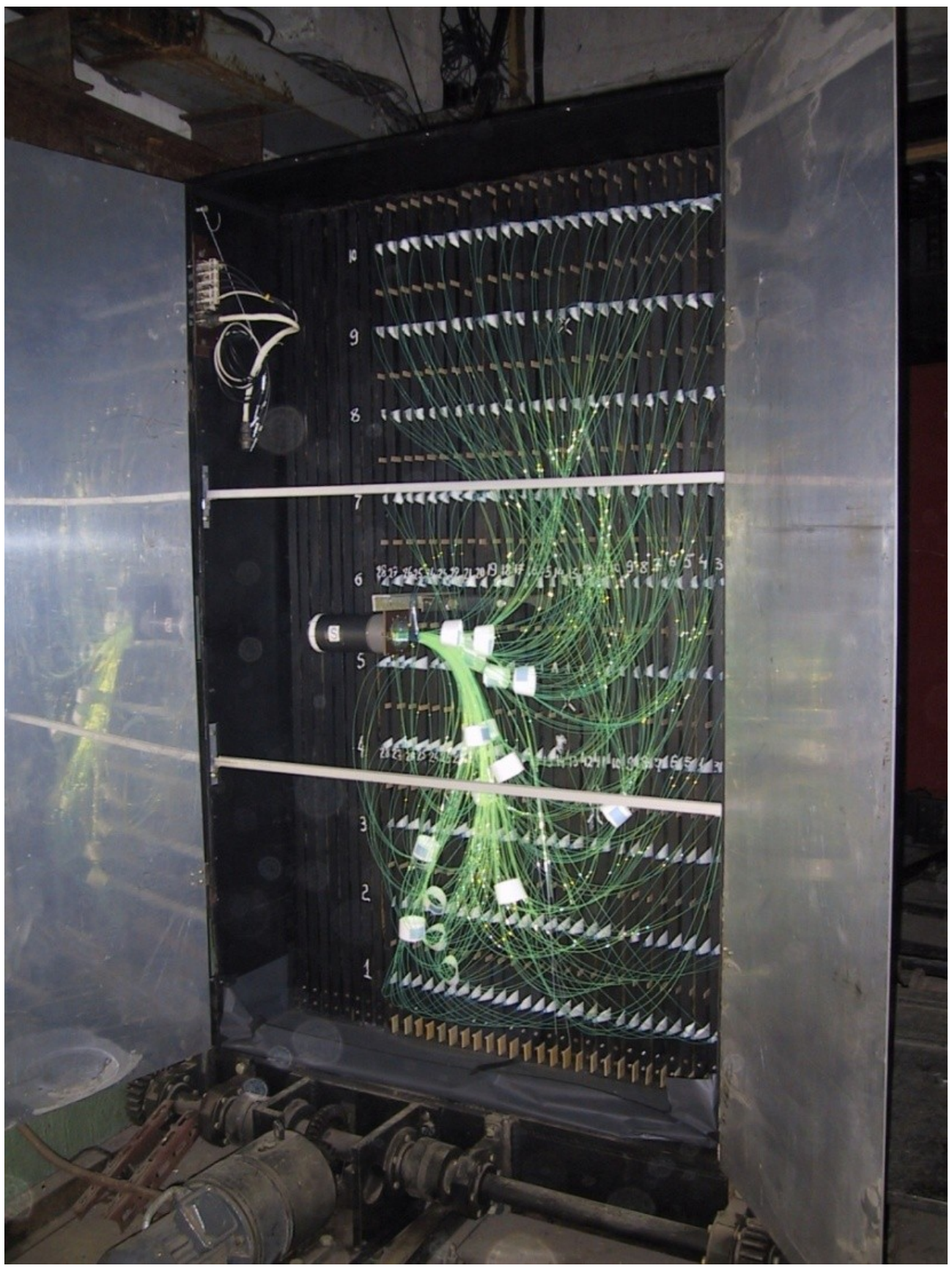

Fig. 6. Photo of the hadron calorimeter with open covers.

## 2. Calorimeter calibration

The calibration was carried out with a secondary negative particles (10, 20, and 30 GeV energy), protons (50 GeV) and electrons (10 and 20 GeV). The calorimeter was placed on the beam axis (fig. 7) and the spectrometer magnet was rotated in such a way that the beam was passing through the aperture of one arm. The coincidence of the counters $S_1S_2S_3\,S_4$ were used for pions and electrons and the counter coincidence $S_1S_2S_3S_4\,S_5$ was used to select muons.

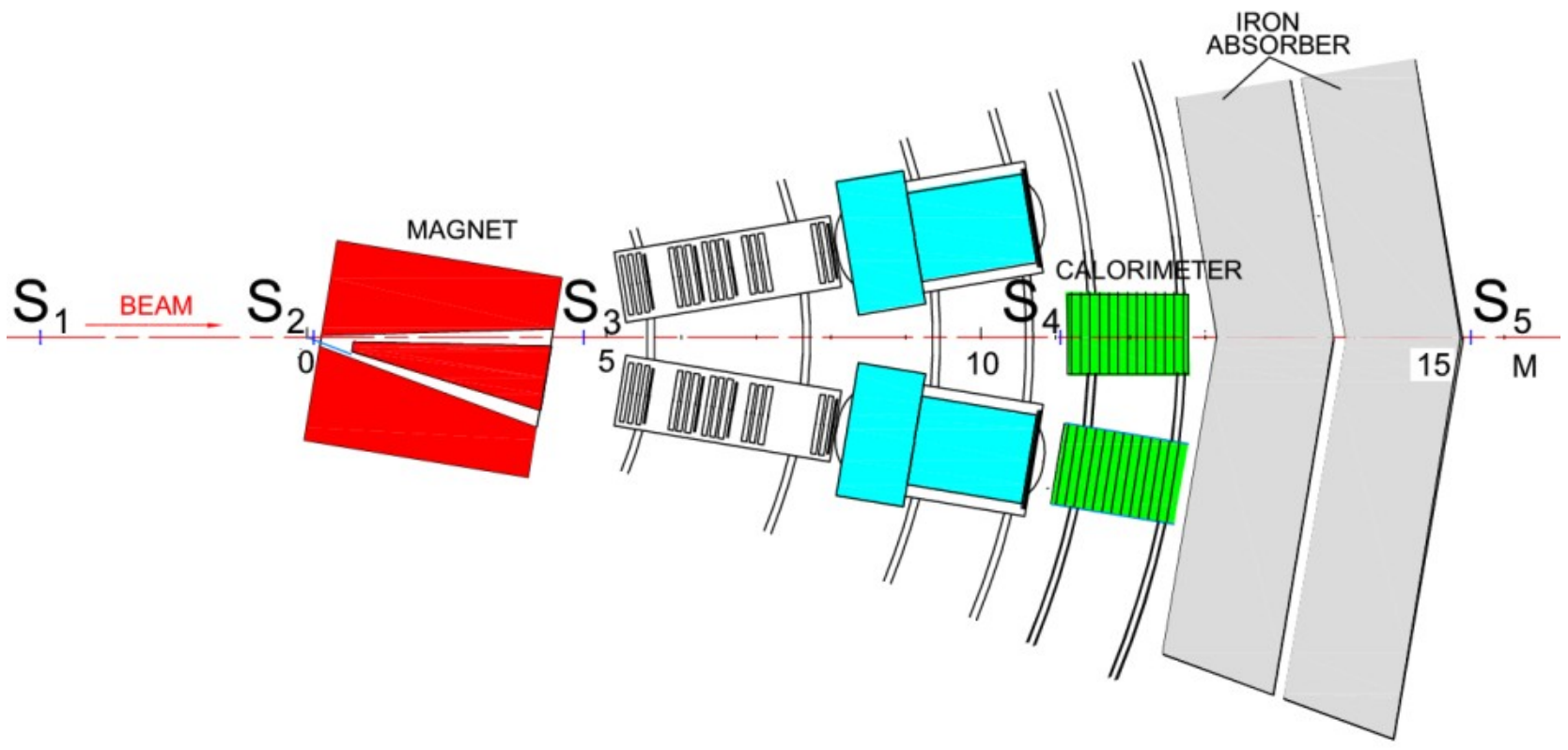

Fig. 7. The layout of the calorimeter calibration. $S_i$ are the scintillation counters.

The phototube signal was fed to ADC. The signal integration time was 100 ns. Fig. 8a. shows the pulse height distribution for muon trigger. Fitting the left side of the distribution by Gauss the number of the photoelectrons can be estimated. For presented distribution the number is 20.

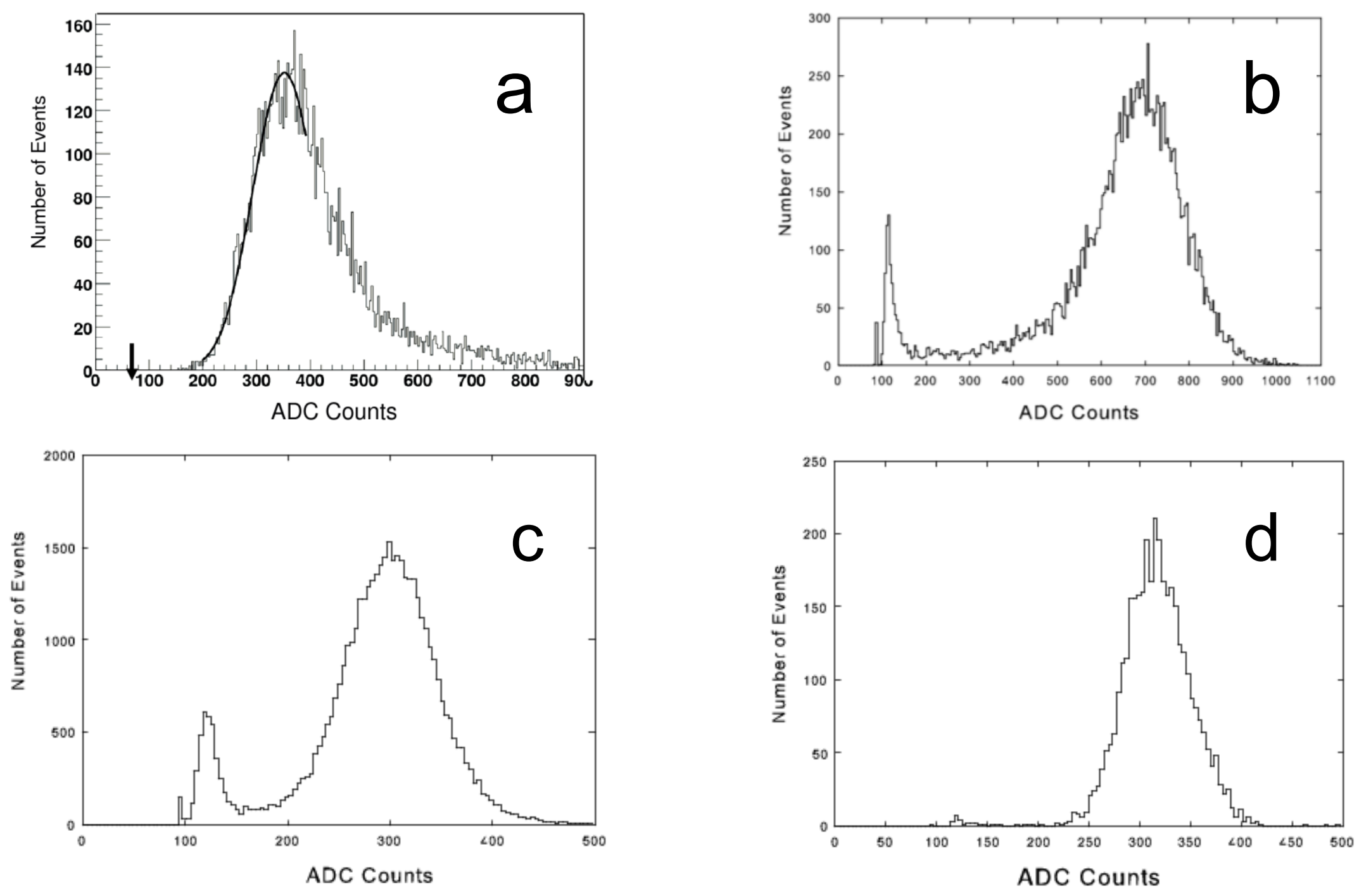

Fig. 8. The calorimeter pulse height distributions: a) with the muon trigger, the pedestal is marked with arrow; b) and c) for secondary particles at 30 and 10 GeV/c momenta correspondingly; d) for 10 GeV/c momentum electrons.

Figs. 8b, 8c, and 8d depict pulse height distributions for pion and electron beams (with muon contamination). At 10 GeV/c hadron momentum the energy resolution is 21.5% that is close to the one defined by a sampling. At the higher hadron energy the asymmetry of the pulse height distribution is increasing due to the energy leakage. For the electron beam the pulse height distribution is symmetrical one and the energy resolution is 14% and 10% at 10 GeV and 20 GeV correspondingly. The maximum pulse height distribution for muon corresponds to 1.32 GeV. The ratio e/h for the calorimeter is 1.04.

The fig 9 presents the dependence of the calorimeter pulse height amplitude on the pion energy. The vertical axis shows the maximum amplitude for hadrons. At 50 GeV there is clear deviation from the linearity.

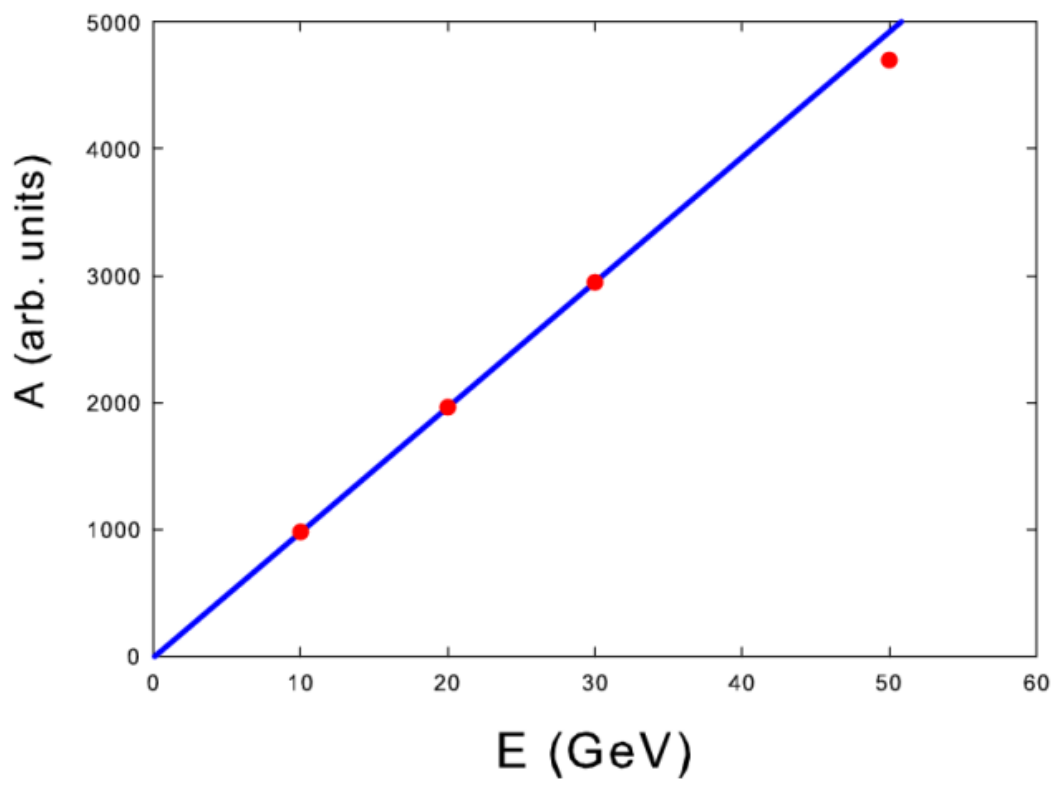

F ig. 9. Dependence of the calorimeter pulse height amplitude on the pion energy.

**Conclusion**

To make a trigger on hadron with the energy above some threshold the hadron calorimeters were upgraded. The design based on the light collection from scintillators by wave shifting fibers to one phototube is simple, easy to calibrate and provides a fast trigger.